\documentclass[onecolumn]{revtex4-1}
\usepackage{amsmath}
\usepackage{amssymb,mathrsfs,eufrak}
\usepackage{graphics}
\usepackage{empheq}
\usepackage{graphicx,epsfig,epsf,color,hhline}

\newcommand\so{\hat \sigma}

\newcommand\Ec{\mathcal{E}}
\newcommand\Fc{\mathcal{F}}
\newcommand\Gc{\mathcal{G}}

\newcommand\Mc{\mathcal{M}}
\newcommand\Nc{\mathcal{N}}

\newcommand\Tr{\mathrm{Tr}}

\newcommand\half{\frac{1}{2}}

\begin{document}

\title{Qubit Entanglement generation by Gaussian non-Markovian dynamics}

\author{F. Benatti}
\email{benatti@ts.infn.it}
\affiliation{Department of Physics, University of Trieste, Strada costiera 11, 34151 Trieste, Italy\\
INFN, Sezione di Trieste, Strada costiera 11, 34151 Trieste, Italy}  
\author{L. Ferialdi}
\email{l.ferialdi@soton.ac.uk}
\affiliation{Department of Physics, University of Trieste, Strada costiera 11, 34151 Trieste, Italy\\
INFN, Sezione di Trieste, Strada costiera 11, 34151 Trieste, Italy}
\affiliation{Department of Physics and Astronomy, University of Southampton, SO17 1BJ, United Kingdom}
\author{S. Marcantoni}
\email{stefano.marcantoni@ts.infn.it}
\affiliation{Department of Physics, University of Trieste, Strada costiera 11, 34151 Treste, Italy\\
INFN, Sezione di Trieste, Strada costiera 11, 34151 Trieste, Italy}


\begin{abstract}
We consider two qubits interacting with a common bosonic bath, but not directly between themselves. We derive the (bipartite) entanglement generation conditions for  Gaussian non-Markovian dynamical maps and show that they are similar as in the Markovian regime; however, they depend on different physical coefficients and hold on different time scales. Indeed, for small times, in the non-Markovian regime entanglement is possibly generated on a shorter time scale ($\propto t^2$) than in the Markovian one ($\propto t$).
Moreover, although the singular coupling limit of non-Markovian dynamics yields Markovian ones, we show that the same limit does not lead from non-Markovian entanglement generation conditions to Markovian ones. 
Also, the entanglement generation conditions do not depend on the initial time  for non-Markovian open  
dynamics resulting from couplings to bosonic Gaussian baths, while  they may depend on time for open dynamics originated by couplings to classical, stochastic Gaussian environments.
\end{abstract}

\maketitle

\section{Introduction}
One of the most interesting features of quantum theory is entanglement. 
It has been the focus of much investigation in the last decades as it plays a crucial role in quantum information theory and its applications, such as quantum computation~\cite{computation} and quantum cryptography~\cite{cryptography}. With them, the interest in understanding how independent, uncorrelated systems may get entangled has been constantly growing~\cite{ent}. Also, many studies have been 
performed to investigate the robustness of entanglement against noise, analyzing the behaviour of correlated quantum systems asymptotically in time~\cite{asymptotic}. 

In this paper we consider a bipartite system consisting of two qubits that interact with a common environment, but not with each other: we shall thus be dealing with open quantum systems, but we will not assume any Markovian approximation.
Indeed, purpose of the paper is to investigate the (bipartite) entanglement generation properties of Gaussian non-Markovian dynamics and to compare them with those holding in a Markovian regime.
Entanglement generation will be identified with the ability of the time-evolution to turn an initial separable two-qubit state into an entangled one~\cite{BenFlo05}. The initial states can be taken to be pure: indeed, if the dynamics is not able to entangle pure separable states, it cannot entangle mixed states either. Furthermore, as we are not interested in the fate of already entangled states, 
we will focus upon the behaviour of pure separable states at small times. In particular, since partial transposition is in this case an exhaustive entanglement witness~\cite{Hor09}, the study can be limited to checking the lack of positive semi-definiteness in the small time expansion of time-evolving two-qubit states.

This kind of analysis has already been performed in the Markovian regime that is for dissipative quantum dynamics obeying a semigroup composition law~\cite{BenFlo05,GKS,Lin76}, where necessary and sufficient conditions for dissipative entanglement generation have been given~\cite{BenFloPia03,BenLigNag08}. However, the behaviour of many open quantum systems is not adequately described by a Markovian dynamics~\cite{physNM}: it is thus natural  and interesting to investigate the issue of entanglement generation in a non-Markovian setting, the results obtained in the Markovian regime thus providing a benchmark.
Precisely, the study of entanglement generation will now be extended to the specific class of Gaussian non-Markovian dynamics: these include models widely used in the description of physical systems, ranging from the spin-boson model~\cite{Legetal87}, quantum Brownian motion~\cite{HPZ} and collapse models~\cite{coll}.  Gaussian non-Markovian dynamics are described by families of maps that can be obtained either from a microscopic, namely quantum, or from an effective, classical stochastic description of the bath~\cite{DioFer14}, both completely characterized by their two-point correlation functions. 

We stress that we call these maps non-Markovian because they do not compose as a two-parameter semigroup. Many different definitions and witnesses of non-Markovianity in the quantum domain have been proposed in the literature and a complete agreement on which one to prefer has not yet been reached. Some of them are based on the distinguishability of quantum states~\cite{BLP}, others rely on the CP-divisibility of dynamical maps~\cite{div}, on the volume of accessible states~\cite{vol}, on the mutual information between system and environment~\cite{mutual}, on the capacity of quantum channels~\cite{capacity}. 
The relation among different definitions of non-Markovian dynamics is an open issue outside the scope of the present work. Some results in this direction can be found in~\cite{comparison}; see also the recent reviews~\cite{nonMarkov}.

In the above physical context, we prove 1)~that non-Markovian entanglement generation occurs, for small times, on a shorter time-scale ($\propto t^2$) than in the Markovian one ($\propto t$);  2)~that, though similar, the entanglement generation conditions in the two regimes involve different physical coefficients, 3) that, the conditions in the Markovian regime cannot be derived as a singular limit of the non-Markovian ones even if the Markovian open dynamics can be derived in this way from the non-Markovian one, 4)~that the non-Markovian entanglement generation conditions do not depend on the initial time for open dynamics derived from coupling to microscopic bosonic baths, while they may vary in time for open dynamics resulting from coupling to classical, stochastic Gaussian fields.

The paper is organized as follows: in Section II, we briefly review the entanglement generation criterion for two qubits in the Markovian regime. In Section III, we introduce a generic Gaussian non-Markovian dynamics for two qubits and discuss the possibility of environment induced entanglement generation at general initial time $t_0$. Finally, Section IV is devoted to the analysis of the differences between Markovian and non-Markovian entanglement generation, while in Section V we draw our conclusions.

\section{Markovian entanglement generation}

We consider two qubits interacting with a common bath but not directly between themselves. Firstly, we assume that their reduced dynamics is Markovian and given by a one-parameter semigroup of completely positive, trace-preserving maps 
$\Gc_t$. Under these conditions, the master equation for the time-evolving two-qubit density matrix $\rho_t:=\Gc_t[\rho]$  is of the typical form~\cite{GKS,Lin76} 
\begin{eqnarray}
&&\partial_t\rho_t=-i[H_S+\widetilde{H}_S,\rho_t]\,+
\label{ME1}
\,\sum_{j,k=1}^3\sum_{\alpha,\gamma=1}^2
K^{\alpha\gamma}_{jk}\Big(\sigma^\alpha_j\,\rho_t\,\sigma^\gamma_k\,-\frac{1}{2}\{\sigma^\gamma_k\sigma^\alpha_j\,,\rho_t\,\}\Big)\ ,
\end{eqnarray}
with initial condition $\rho$. The right hand side of the above expression is the Gorini-Kossakowski-Sudarshan-Lindblad generator $G$ of the reduced dynamics, $\Gc_t=\exp(t\,G)$. 
The second term in $G$ is a purely dissipative contribution due to the presence of the bath, with
 $\sigma^{\alpha}_j$, $\alpha=1,2$, the Pauli matrices associated with qubit $\alpha$. Instead, the first term is the commutator with a Hamiltonian,  where $H_S$ is the initial two qubit Hamiltonian without interaction as in Eq.~\eqref{ham} below, and
 \begin{equation}
 \widetilde{H}_S= \sum_{j,k=1}^3\sum_{\alpha,\gamma=1}^2 h^{\alpha\gamma}_{jk}\sigma^\alpha_j \sigma^\gamma_k,
 \end{equation}
  is a dynamical coupling mediated by the bath, resulting from Markovian approximations like the so called weak coupling limit~\cite{BrePet02}.

The Kossakowski coefficients $K^{\alpha\gamma}_{jk}=\overline{K^{\gamma\alpha}_{kj}}$, where the bar denotes complex conjugation, form a self-adjoint matrix $K$ which must be positive semi-definite in order to guarantee the complete positivity of the maps $\Gc_t$. This matrix can be conveniently written as a $6\times 6$
matrix in qubit blocks,
\begin{equation}
\label{KosMat}
K:=\begin{pmatrix}K^{11}&K^{12}\cr K^{21}&K^{22}
\end{pmatrix}\ ,\qquad K^{21}=(K^{12})^\dag\ , 
\end{equation}
with $K^{\alpha\gamma}$ $3\times 3$ matrices: $K^{11}$ and $K^{22}$ are positive semi-definite and pertain to the two qubits independently, while $K^{12}$ and $K^{21}$ statistically correlate them.\\ 
 For the sake of comparison with the non-Markovian time evolution that will be introduced in Eq.~\eqref{GMtt0b}, we will now pass to the interaction representation with respect to the Hamiltonian $H_S$. The formal integration of the master equation~\eqref{ME1} then gives
a dynamical map $\Fc_{t,t_0}[\rho]=\rho_t$ of the form
\begin{eqnarray}
\nonumber
\Fc_{t,t_0}&=&\mathbb{T}\exp\Bigg\{\int_{t_0}^td\tau \sum_{j,k,\alpha,\gamma} \bigg[K^{\alpha\gamma}_{jk}\bigg(\so^{\alpha}_{jL}(\tau-t_0)\so^{\gamma}_{kR}(\tau-t_0)
-\frac{1}{2}\so^{\gamma}_{kL}(\tau-t_0)\so^{\alpha}_{jL}(\tau-t_0)
\label{GMtL}
-\frac{1}{2}\so^{\alpha}_{jR}(\tau-t_0)\so^{\gamma}_{kR}(\tau-t_0)\bigg)\\
&&\hspace{3.5cm}-ih^{\alpha\gamma}_{jk} \bigg(\sigma^\alpha_{jL}(\tau-t_0)\sigma^\gamma_{kL}(\tau-t_0)-\sigma^\gamma_{kR}(\tau-t_0)\sigma^\alpha_{jR}(\tau-t_0)\bigg)\bigg]\Bigg\}\nonumber\\
\label{F}\hskip .8cm
&=&\mathbb{T}\exp\Bigg\{\!\!\int_{0}^{t-t_0}\hspace{-0.5cm}d\tau \!\!\sum_{j,k,\alpha,\gamma}\!\!\bigg[ K^{\alpha\gamma}_{jk}\bigg(\!\so^{\alpha}_{jL}(\tau)\so^{\gamma}_{kR}(\tau)
\label{GMtL2}
\!-\!\frac{1}{2}\so^{\gamma}_{kL}(\tau)\so^{\alpha}_{jL}(\tau)\!-\!\frac{1}{2}\so^{\alpha}_{jR}(\tau)\so^{\gamma}_{kR}(\tau)\!\!\bigg)\!-\!i\widetilde{H}_{SL}(\tau)\!+i\widetilde{H}_{SR}(\tau)\!\bigg]=:\Fc_{t-t_0}\,.
\end{eqnarray}
In the above expression, $\mathbb{T}$ is the time ordering operator and we used the following short-hand notation for the left and right operator multiplication (note that superoperators are denoted with hats, while operators are not):
\begin{equation}
\label{notation1}
\so^{\alpha}_{jL}(\tau)\,\rho=\sigma^{\alpha}_j(\tau)\,\rho\ ,\quad \so^{\gamma}_{kR}(\tau)\,\rho=\rho\,\sigma^{\gamma}_k(\tau)
\end{equation}
with
\begin{equation}
\label{notation2}
\sigma^{\alpha}_j(\tau)={\rm e}^{i H_S\,\tau}\sigma^{\alpha}_j{\rm e}^{-iH_S\,\tau}\ .
\end{equation}
The dynamical maps 
$\Fc_{t,t_0}$ are thus time homogeneous, $\Fc_{t,t_0}=\Fc_{t-t_0}$, and form a two-parameter semigroup, {\it i.e.} they satisfy the composition law $\Fc_{t,t_0}=\Fc_{t,s}\circ\Fc_{s,t_0}$ for all $t\geq s\geq t_0$, and 
thus describe a regime commonly agreed to call Markovian~\cite{GKS,Lin76}. Such a property has relevant consequences for entanglement generation since, as we shall see, it then becomes sufficient to investigate the dynamics around time $t_0=0$. We shall thus expand the action of $\Fc_t$ as follows:
\begin{equation}\label{expansL}
\Fc_t\simeq\Fc_{t=0}+t\,\Fc'_{t=0}+\dots\,.
\end{equation}
Then, the short time expansion of the two-qubit density matrix $\rho_t$ about the initial condition $\rho$ is
\begin{eqnarray}
\hskip-.2cm
&&\rho_t\simeq\rho+
\label{shortM}
t\,\sum_{j,k=1}^3\sum_{\alpha,\gamma=1}^2
\left(K^{\alpha\gamma}_{jk}\Big(\sigma^{\alpha}_j\,\rho\,\sigma^\gamma_k\,-\,\frac{1}{2}\Big\{
\sigma^\gamma_k\sigma^\alpha_j\,,\,\rho\Big\}\Big)-i h^{\alpha\gamma}_{jk}\left[\sigma^\alpha_j \sigma^\gamma_k,\rho\right]\right)\ .
\end{eqnarray}
We stress that, as expected, both Hamiltonian and dissipative terms of Eq.~\eqref{ME1} contribute to the short time evolution of the dynamics.
In the following we are interested in the entanglement generation properties of the dynamics. Since the free $H_S$ does not contain interactions between the two qubits, such properties can be studied in the chosen interaction representation, without need of going back to the physical representation. Indeed, entanglement generation can only come from the bath contributed terms in the generator $G$.
We now exploit the fact that, for two qubits, partial transposition is an exhaustive entanglement witness~\cite{Hor09}. By performing this operation on the second qubit (${\rm id}\otimes T$), and setting $\rho^*_t={\rm id}\otimes T[\rho_t]$,  we find
\begin{eqnarray}
&&\rho^*_t\simeq\rho^*\,+\,t\sum_{j,k=1}^3\Bigg[ K^{11}_{jk} \left(\sigma^{1}_j\,\rho^*\,\sigma^{1}_k-\half\{\sigma_k^1\sigma_j^1\,,\,\rho^*\}\right)-\epsilon_k\,\left(\Re(K^{12}_{jk})+ih^{12}_{jk}\right) \left(\sigma^{1}_j\,\rho^*\,\sigma^{2}_k-\half\{\sigma_k^2\sigma_j^1\,,\,\rho^*\}\right) \nonumber\\
&&\hspace{3cm}-\epsilon_j\,\left(\Re(K^{12}_{kj})-ih^{12}_{kj}\right) \left(\sigma^{2}_j\,\rho^*\,\sigma^{1}_k-\half\{\sigma_k^1\sigma_j^2\,,\,\rho^*\}\right)+\epsilon_j\,\epsilon_k\,K^{22}_{kj} \left(\sigma^{2}_j\,\rho^*\,\sigma^{2}_k-\half\{\sigma_k^2\sigma_j^2\,,\,\rho^*\}\right)  \nonumber \\
&&\hspace{3cm}-i h^{11}_{jk} \Big[ \sigma^{1}_j  \sigma^{1}_k, \rho^*  \Big]+ i \epsilon_j \,\epsilon_k \,  h^{22}_{jk} \Big[  \sigma^{2}_k \sigma^{2}_j , \rho^*  \Big]+\Im(K^{12}_{jk})\,\epsilon_k\,\Big[\sigma^1_j\sigma^2_k\,,\,\rho^*\Big]
\Bigg]\ ,\nonumber
\label{ptME}
\end{eqnarray}
where $\Re$ and $\Im$ denote real and imaginary parts, while the factors $\epsilon_{1,3}=1$, $\epsilon_{2}=-1$ come from choosing the standard representation of the Pauli matrices, so that, under transposition, $\sigma_1^T=\sigma_1$, $\sigma_3^T=\sigma_3$, while $\sigma_2^T=-\sigma_2$. One has now to check whether $\rho_t^*$ is positive semi-definite or not: in the latter case $\rho_t$ is entangled.
Notice that partial transposition yields an expression similar to~\eqref{shortM}, with the Kossakowski matrix $K$ in~\eqref{KosMat} replaced by
\begin{equation}
\label{ptKosMat}
\widetilde{K}:=\begin{pmatrix}1&0\cr0&\Ec\end{pmatrix}
\begin{pmatrix}
K^{11}&\Re(K^{12})+ih^{12}\cr
(\Re(K^{12})-ih^{12})^T&(K^{22})^T
\end{pmatrix}\begin{pmatrix}1&0\cr0&\Ec\end{pmatrix}
\ ,
\end{equation}
with $\mathcal{E}:=\hbox{diag}(-1,1,-1)$ and $\Re(K^{12})$ the $3\times 3$ matrix with entries given by the real parts of the entries of $K^{12}$, $(K^{12})^T$ denoting its transposed. As much as the matrix $K$ of the generator $G$ in~\eqref{ME1}, the matrix $\widetilde{K}$ is connected with the generator $\widetilde{G}$ of a semigroup of maps 
$\widetilde{\Gc}_t={\rm id}\otimes T\circ\Gc_t\circ{\rm id}\otimes T\ ,\ t\geq 0$,
such that $\rho^*_t=\widetilde{\Gc}_t[\rho^*]$.
However, unlike $K$, $\widetilde{K}$ 
need not be positive semidefinite. 
Should $\widetilde{K}$ result positive semi-definite, the maps $\widetilde{\Gc}_t$ would then be completely positive and thus preserve the positivity of any initial two-qubit state, namely also that of separable states $\rho=\vert\psi\rangle\langle\psi\vert\otimes \vert\phi\rangle\langle\phi\vert$. In such a case, due to the semigroup property,  entanglement could not be generated either at small or at any other time $t\geq 0$. On the other hand, if $\widetilde{K}$ is not positive semi-definite the conditions  
for entanglement generation have already been extensively investigated~\cite{BenFloPia03,BenLigNag08}. One finds that $\Gc_t$ is able to entangle $\vert\psi\rangle\langle\psi\vert\otimes \vert\phi\rangle\langle\phi\vert$, at first order in $t>0$, if and only if
\begin{equation}\label{entgenM}
\langle u | K^{11}| u\rangle\langle v | (K^{22})^T| v\rangle-|\langle u | \Re(K^{12})+ih^{12}| v\rangle|^2<0\ ,
\end{equation}
where 
\begin{equation}
\label{vec}
\vert u\rangle=(\langle\psi\vert\sigma_i\vert\psi_\perp\rangle)_{i=1}^3\ ,\quad 
\vert v\rangle=(\langle\phi_\perp\vert\sigma_i\vert\phi\rangle)_{i=1}^3\ ,
\end{equation}
with $\vert\psi\rangle,\vert\psi_\perp\rangle$ and $\vert\phi\rangle,\vert\phi_\perp\rangle$ two orthonormal bases in the qubit Hilbert space $\mathbb{C}^2$.

In the non-Markovian regime, an open quantum dynamics is generated by a master equation of the form
\begin{eqnarray}
\hskip-.5cm
\partial_t\rho_t&=&-i[H_S+\widetilde{H}_S(t),\rho_t]\,+
\label{NME1}
\sum_{j,k=1}^3\sum_{\alpha,\gamma=1}^2
K^{\alpha\gamma}_{jk}(t)\Big(\sigma^\alpha_j\,\rho_t\,\sigma^\gamma_k\,-\frac{1}{2}\{\sigma^\gamma_k\sigma^\alpha_j\,,\rho_t\,\}\Big)
\end{eqnarray}
with explicitly time-dependent Hamiltonian $\widetilde{H}_S(t)$ and coefficients $K^{\alpha\gamma}_{jk}(t)$ which form a hermitian, but not positive semi-definite matrix. 
The above expressions are the most general non-Markovian master equations under the assumption that the generated dynamical maps $\Lambda_t$ posses an inverse
$\Lambda^{-1}_t$: in such cases, as proved in~\cite{ChruKos}, also time non-local master equations of the form 
$$
\partial_t\rho_t=\int_0^t {\rm d}\tau\, \mathbb{K}(t-\tau)\rho_\tau\ ,
$$
with a given operatorial kernel $\mathbb{K}(t)$, can be turned into  time local ones 
$$
\partial_t\rho_t=\mathbb{L}_t[\rho_t]\ .
$$
Furthermore, the latter can always be written as in~\eqref{NME1}, however with in general a non-positive matrix $[K^{\alpha\gamma}_{jk}(t)]$.

Because of the explicit time-dependence, one 
expects that the entanglement generation conditions differ from those in the Markovian case and themselves explicitly depend on time.
In the following, we approach this issue by means of the environment two-point correlation functions upon which the coefficients $K^{\alpha\gamma}_{jk}(t)$ depend and show 
that different behaviours emerge depending on whether the environment is a bosonic thermal bath, or consists of classical, Gaussian stochastic fields.

\section{Non-Markovian entanglement generation}

Unlike Markovian dynamics, non-Markovian ones cannot in general be recast in a unique form. 
However, a complete characterization has recently been achieved for the class of Gaussian non-Markovian dynamics~\cite{DioFer14}. 
These dynamics can be obtained either by a microscopic description of the environment by means of non-commuting bosonic fields, or by an effective description of the environment based on commuting, that is classical, 
stochastic Gaussian fields. It turns out that there are dissipative time-evolutions resulting from the stochastic approach that cannot be obtained from the microscopic one~\cite{Bud15}.

 We consider a model described by a  Hamiltonian 
\begin{equation}
\label{ham}
H=H_S+H_B+H_I\ ,\ H_S=\sum_{\alpha,j} \omega^{\alpha}_j \sigma^{\alpha}_j\,,
\end{equation}
and we distinguish two cases.
In the case of a derivation based on a microscopic description of the environment, one assumes that the system is bilinearly interacting with a bath of independent bosons:
\begin{equation}
H_B=\sum_j\omega_jb^{\dag}_jb_j\ ,\ H_I=\sum_{\alpha, j} \sigma^{\alpha}_j\phi^{\alpha}_j
\ ,
\label{ham1}
\end{equation}
where the fields $\phi^\alpha_j$ are hermitian linear combinations of the bath creation and annihilation operators such that $[b_j\,,\,b^\dag_k]=\delta_{jk}$. The environment  state $\rho_B$ is then assumed to be central Gaussian that assigns  zero mean values to the fields $\phi^\alpha_j$ and  is completely characterized by two-point correlation functions
\begin{equation}
\label{boscorr}
D_{jk}^{\alpha\gamma}(\tau,s):=\Tr_B[\phi^{\gamma}_k(s) \phi^{\alpha}_j(\tau) \rho_B]\ ,
\end{equation} 
where $\rho_B$ is not necessarily invariant under $H_B$.

As an alternative approach, one can choose an effective description of the environment, by assuming that the qubits are interacting with (classical) complex stochastic Gaussian fields  $\phi^\alpha_j$ 
completely characterized by the following correlation functions
\begin{eqnarray}\label{corrD}
D_{jk}^{\alpha\gamma}(\tau,s)&=&\mathbb{E}\left[ \phi_j^{\alpha}(\tau)\overline{\phi_k^{\gamma}}(s)\right]\,,\\
\label{corrS}S_{jk}^{\alpha\gamma}(\tau,s)&=&\mathbb{E}\left[\phi_j^{\alpha}(\tau)\,\phi_k^{\gamma}(s)\right]\,,
\end{eqnarray}
where $\mathbb{E}$ denotes the stochastic average. In this case $H_B=0$, and $H_I$ is defined as in~\eqref{ham1}.

If the interaction starts at $t_0$, by generalising the result of~\cite{DioFer14}, one finds that the Gaussian non-Markovian reduced map describing the qubits dynamics reads
\begin{equation}
\label{GMta}
\hskip-.5cm
\Mc_{t,t_0}=\mathbb{T}\exp(M_{t,t_0})\,,
\end{equation}
with
\begin{eqnarray}
M_{t,t_0}&=&\int_{t_0}^td\tau\int_{t_0}^{\tau}ds \sum_{j,k,\alpha,\gamma} \left[\so^{\alpha}_{jL}(\tau-t_0)-\so^{\alpha}_{jR}(\tau-t_0)\right]
\label{GMtt0b}
\left[\mathfrak{D}_{jk}^{\alpha\gamma}(\tau,s;t_0)\so^{\gamma}_{kR}(s-t_0)-\overline{\mathfrak{D}_{jk}^{\alpha\gamma}(\tau,s;t_0)}\so^{\gamma}_{kL}(s-t_0)\right]\,.
\end{eqnarray}
The time dependence of the actions of the $\so$ is a direct consequence of Eq.~\eqref{notation2} when the initial time is $t_0>0$.
The two-point correlation function is defined as follows: $\mathfrak{D}_{jk}^{\alpha\gamma}(\tau,s;t_0):=D_{jk}^{\alpha\gamma}(\tau-t_0,s-t_0)$ if one considers a microscopic description; $\mathfrak{D}_{jk}^{\alpha\gamma}(\tau,s;t_0):=D_{jk}^{\alpha\gamma}(\tau,s)$ for the stochastic case. One can easily check that $\overline{\mathfrak{D}_{jk}^{\alpha\gamma}(\tau,s;t_0)}=\mathfrak{D}_{kj}^{\gamma\alpha}(s,\tau;t_0)$, and that the matrix $[\mathfrak{D}^{\alpha\gamma}_{jk}(t,t;t_0)]$ is positive semi-definite for all $t,t_0$.  One can rearrange the terms of Eq.~\eqref{GMtt0b} in order to write it in a form closer to Eq.~\eqref{F}, by separating a purely dissipative contribution, with coefficients $\mathfrak{D}_{jk}^{\alpha\gamma}(\tau,s;t_0)$, from a commutator with respect to a Hamiltonian of the form
\begin{equation}\label{hamburger}
\widetilde{H}_S(\tau,s;t_0)= \frac{1}{2}\sum_{j,k,\alpha,\gamma}\left[-i\Re\left(\mathfrak{D}_{jk}^{\alpha\gamma}(\tau,s;t_0)\right)\left[\so^{\alpha}_{j}(\tau-t_0),\so^{\gamma}_{k}(s-t_0)\right]-\Im\left(\mathfrak{D}_{jk}^{\alpha\gamma}(\tau,s;t_0)\right)\left\{\so^{\alpha}_{j}(\tau-t_0),\so^{\gamma}_{k}(s-t_0)\right\}\right]
\end{equation}
We stress that the main difference between Eq.~\eqref{F} and Eq.~\eqref{GMtt0b} is that, while the first one displays a single integration in time, the latter one presents two integrations. Accordingly, as a direct consequence of the properties of time-ordering $\mathbb{T}$, $\Fc_{t,t_0}$ satisfies the two-parameter semigroup property, but $\Mc_{t,t_0}$ does not: the lack of such a time-composition law is what we mean by non-Markovian maps. Interestingly, a Markov limit of $\Mc_{t,t_0}$ is simply obtained by choosing 
singular two-point correlation functions of the type $\mathfrak{D}_{jk}^{\alpha\gamma}(t-s)=K_{jk}^{\alpha\gamma}\delta(t-s)$, corresponding to an uncorrelated bath: then, direct substitution  
in Eq.~\eqref{GMtt0b} leads to a Markovian map of the type~\eqref{GMtL}.

In order to investigate the entanglement generation properties of $\Mc_{t,t_0}$, we adopt the same strategy as in the Markovian case.  We expand it in Taylor series:
\begin{equation}
\label{expansNM}
\Mc_t\simeq\Mc_{t=t_0}+(t-t_0)\,\Mc'_{t=t_0}+\frac{(t-t_0)^2}{2}\, \Mc''_{t=t_0}+\dots\,.
\end{equation}
One finds that $\Mc_{t=t_0}=\mathbb{I}$ and  $\Mc'_{t=t_0}=0$, while
\begin{equation}
\label{expansNM1}
\Mc''_{t=t_0}=\sum_{j,k,\alpha,\gamma}(\so^{\alpha}_{jL}-\so^{\alpha}_{jR})\left(D_{jk}^{\alpha\gamma}\so^{\gamma}_{kR}- \overline{D_{jk}^{\alpha\gamma}}\so^{\gamma}_{kL}\right)\,,
\end{equation}
where we set $D^{\alpha\gamma}_{jk}	:= \mathfrak{D}^{\alpha\gamma}_{jk}(t_0,t_0;t_0)$.
Therefore, the following short time behaviour holds:
\begin{eqnarray}
&&\hskip-.5cm
\rho_t\simeq\rho_{t_0}\,+\frac{(t-t_0)^2}{2}\sum_{j,k=1}^3\sum_{\alpha,\gamma=1}^2 D_{jk}^{\alpha\gamma}\Big(\sigma^{\alpha}_j\,\rho_{t_0}\,\sigma^\gamma_k\,-\,\frac{1}{2}\{\sigma^\gamma_k\sigma^\alpha_j,\rho_{t_0}\}\Big)\ .
\label{2order}
\end{eqnarray}
 Remarkably, unlike the Markovian regime, in the non-Markovian one the Hamiltonian terms do not contribute to entanglement generation. Indeed, the property $\overline{\mathfrak{D}_{jk}^{\alpha\gamma}(\tau,s;t_0)}=\mathfrak{D}_{kj}^{\gamma\alpha}(s,\tau;t_0)$ implies that $\Re(D_{jk}^{\alpha\gamma})= \Re(D_{kj}^{\gamma\alpha})$, and $\Im(D_{jk}^{\alpha\gamma})= -\Im(D_{kj}^{\gamma\alpha})$. Exploiting these properties in Eq.~\eqref{hamburger}, one easily finds that $\widetilde{H}_S(t_0,t_0;t_0)=0$. This is a first hint of the fact that the small time expansion, performed to obtain Eq.~\eqref{shortM} and Eq.~\eqref{2order}, and the Markov limit do not commute. This issue will be further discussed in Sec.IV.
Concerning the dissipative contributions of Eq.~\eqref{2order}, besides the quadratic dependence on time, 
the small time expansion in the non-Markovian regime has the same structure as the small time expansion in the Markovian regime in Eq.~\eqref{shortM}.
The only difference is that the Kossakowski matrix with entries $K_{jk}^{\alpha\gamma}$ proper of the Markovian regime is replaced by the matrix $D$ with entries $D_{jk}^{\alpha\gamma}$.
As already observed after Eq.~\eqref{GMtt0b}, the matrix $D$ is automatically positive semi-definite. Then, one can use the Markovian techniques of the previous section to derive the following small time entanglement generation criterion for the non-Markovian Gaussian dynamics:
\begin{equation}
\label{entgenNM}
\langle u | D^{11}| u\rangle\langle v | (D^{22})^T| v\rangle-|\langle u | \Re(D^{12})| v\rangle|^2<0\ ,
\end{equation}
where $D^{\alpha\gamma}$ are the $3\times 3$ matrices with entries $D^{\alpha\gamma}_{jk}$.

Besides showing a coefficient matrix $D=[D^{\alpha\gamma}_{jk}]$ different from the $K=[K^{\alpha\gamma}_{jk}]$ of the Markovian regime, the non-Markovian regime presents a small time expansion that starts with $(t-t_0)^2$ instead of $t-t_0$: this is due to the fact that, in the Markovian limit, the singular (delta) correlation function eliminates one of the integrals in the exponent of $\Mc_t$. This different time-dependence is an interesting feature that allows to discriminate Gaussian non-Markovian dynamics from Markovian ones. 
Indeed, for small times, in the non-Markovian regime, entanglement is generated on a shorter time scale ($\propto t^2$) than in the Markovian one. This issue will be further discussed in the next section.

Some comments are in order at this point. In the Markovian case, if entanglement generation occurs at time $t=t_0$, it then occurs at any later time $t>t_0$. This fact is due to the semigroup composition law, but need not be true if the  
the law fails as in the non-Markovian case. 
In such a case, the criterion~\eqref{entgenNM} can only establish whether the map $\Mc_t$ is able to entangle factorized states at the second order in time around $t=t_0$.  In order to infer entanglement generation at generic later times $t_1>t_0$, after the initial state $\rho_{t_0}$ has evolved into $\rho_{t_1}=\Mc_{t_1,t_0}\rho_{t_0}$, one should explicitly study the interpolating map $\Nc_{t,t_1}$, formally given by
\begin{equation}
\label{tdep}
\Mc_{t,t_0}=\Nc_{t,t_1}\circ \Mc_{t_1,t_0}\ ,\ \Nc_{t,t_1}:=\Mc_{t,t_0}\circ\Mc^{-1}_{t_1,t_0}\ ,
\end{equation}
which links the dynamics from $t=t_0$ to $t>t_0$ to that from $t_1$ to $t>t_1$. 
Unfortunately, apart from very simple systems~\cite{Legetal87,Gar97}, $\Nc_{t,t_1}$ is not available in analytic form (see \textit{e.g.} the discussion in~\cite{Fer17}), also preventing us from comparing our definition of non-Markovian dynamics with that relying on CP-divisibility~\cite{div}.


\subsection{Microscopic description}
As we have previously seen, if one considers a microscopic description the map $\Mc_{t,t_0}$ depends on the environmental correlation functions 
\begin{equation}\label{boscorrt0}
\mathfrak{D}_{jk}^{\alpha\gamma}(\tau,s;t_0):=D_{jk}^{\alpha\gamma}(\tau-t_0,s-t_0)\,.
\end{equation}
This kind of correlation functions is obtained by means of Gaussian initial states that are not invariant under the environment time-evolution. On the other side, if $[H_B,\rho_B]=0$, the correlation functions become time-translation invatiant: $D^{\alpha\gamma}_{jk}(\tau-s)$, loosing their dependence on the initial time $t_0$.
With correlation functions of the type~\eqref{boscorrt0} one can perform a change of variables in the integrals of Eq.~\eqref{GMtt0b} and obtain $\Mc_{t,t_0}=\Mc_{t-t_0,0}$. Accordingly, the ability of generating entanglement of $\Mc_{t,t_0}$ for small times $t-t_0$ is the same as that of $\Mc_{t,0}$ for small $t>0$, i.e. the entanglement generation criterion~\eqref{entgenNM} for this dynamics does not depend on $t_0$.
However, as will be shown in the next section, the specific time dependence of Eq.~\eqref{boscorrt0} need not be satisfied if one considers a stochastic derivation of $\Mc_{t,t_0}$ which indeed allows for more general non time-homogeneous non-Markovian dynamics such that $\Mc_{t,t_0}\,\neq\,\Mc_{t-t_0,0}$.

\subsubsection{Examples}
In order to illustrate which environment correlation  functions  lead to entanglement generation in the case of a microscopic derivation, we assume that the qubits couple to  each environment degree of freedom with different strength, i.e. that the bath coupling operators have the following form:
\begin{equation}
\label{fields}
\phi^{\alpha}_j=\sum_\ell\Big(c_{j\ell}^{\alpha}\,b_\ell+\overline{c_{j\ell}^{\alpha}}\,b_\ell^{\dag}\Big)\,,
\end{equation}
where $c_{jl}^{\alpha}$ are arbitrary complex constants. Note that this is the most general linear bath operator one can choose.
Let us now consider the case of a microscopic derivation of the open system dynamics based on an environment consisting of bosonic fields, $[b_j\,,\,b^\dag_k]=\delta_{jk}$, in a
thermal state at inverse temperature $\beta$ and invariant under its own free dynamics $H_B=\sum_j\omega_jb^{\dag}_jb_j$. One finds the following time-translation invariant two-point correlation functions:
\begin{eqnarray}
\nonumber
&&D_{jk}^{\alpha\gamma}(t-s)=\int d\omega\Big\{J^{\alpha\gamma}_{jk}(\omega)\,{\rm e}^{-i\omega(t-s)}\,\frac{{\rm e}^{\beta\omega}}{{\rm e}^{\beta\omega}-1}+\overline{J^{\alpha\gamma}_{jk}(\omega)}\,{\rm e}^{i\omega(t-s)}\,\frac{1}{{\rm e}^{\beta\omega}-1}\Big\}\,,
\label{Dgen}
\end{eqnarray}
where we have introduced the spectral densities: 
\begin{equation}
J^{\alpha\gamma}_{jk}(\omega):=\sum_\ell \overline{c_{j\ell}^{\alpha}}\, c_{k\ell}^{\gamma}\,\delta(\omega-\omega_\ell)\,.
\end{equation}

We stress that any set of generic time-translation invariant correlation functions $D_{jk}^{\alpha\gamma}(t-s)$ can be derived from suitable microscopic bosonic baths in thermal equilibrium with respect to $H_B$~\cite{DioFer14}. Notice that this is not true for not time-translation invariant two-point correlation functions $D_{jk}^{\alpha\gamma}(t,s)\neq D_{jk}^{\alpha\gamma}(t-s)$: these ones can be obtained either by coupling the qubits to a Gaussian environment state not invariant under $H_B$ (see Eq.~\eqref{ham} and Eq.~\eqref{ham1}) or, as we shall see in the next Section, by coupling them to commuting and stochastic Gaussian fields.

Instead, when $D_{jk}^{\alpha\gamma}(t,s)=D_{jk}^{\alpha\gamma}(t-s)$, as discussed before, the entanglement criterion~\eqref{entgenNM} does not depend on the initial time and  
is based on the entries
\begin{eqnarray}
\nonumber
\hskip-.5cm
D_{jk}^{\alpha\gamma}&=&\sum_\ell\Big\{\overline{c^{\alpha}_{j\ell}} c^{\gamma}_{k\ell} \frac{{\rm e}^{\beta\omega_\ell}}{{\rm e}^{\beta\omega_\ell}-1}+ c^{\alpha}_{j\ell} \overline{c^{\gamma}_{k\ell}}
\frac{1}{{\rm e}^{\beta\omega_\ell}-1}\Big\}\\
\label{mat0}
\hskip-.5cm
&=&\sum_\ell\Big\{\Re(\overline{c^\alpha_{j\ell}}c^\gamma_{k\ell})\,\coth\left(\frac{\beta\omega_\ell}{2}\right)+i\,\Im(\overline{c^\alpha_{j\ell}}c^\gamma_{k\ell})\Big\}\,.
\end{eqnarray}
These, as required, make for a positive semi-definite matrix $D$. Let us consider a few cases.\\

\noindent
$1.$\quad
The coupling constants $c^\alpha_{j\ell}$ are such that $\Re(D^{12}_{jk})=0$: then,
the analog of the matrix $\widetilde{K}$ in Eq.~\eqref{ptKosMat}
with $D^{\alpha\gamma}$ in the place of $K^{\alpha\gamma}$ reads
\begin{equation}
\label{D1}
\widetilde{D}=
\begin{pmatrix}
1&0\cr 0&\mathcal{E}
\end{pmatrix}
\begin{pmatrix}D^{11}&0\cr
0&(D^{22})^T\end{pmatrix}
\begin{pmatrix}1&0\cr0&\mathcal{E}\end{pmatrix}\ .
\end{equation}
Since $D$ is positive semi-definite, so are $D^{11}$, $(D^{22})^T$ and
$\widetilde{D}$. Therefore, the master equation~\eqref{ME1} with Kossakowski matrix given by $\widetilde{D}$ generates completely positive maps $\widetilde{\Gc}_\tau$, with semigroup parameter $\tau=t^2$. These, as argued in the previous section, cannot fulfill the inequality~\eqref{entgenM} and entanglement cannot be generated.\\

\noindent
$2.$\quad The coupling constants are all real or all purely imaginary: then, the $3\times 3$ matrices $D^{11}$ and $D^{22}$ are both symmetric and $\Re(D^{12})=(D^{12}+((D^{12})^\dag)^T)/2$,  so that
\begin{equation}
\label{D2}
\widetilde{D}=
\begin{pmatrix}
1&0\cr 0&\mathcal{E}
\end{pmatrix}\frac{D+D^T}{2}
\begin{pmatrix}1&0\cr0&\mathcal{E}\end{pmatrix}\ .
\end{equation}
Therefore, $\widetilde{D}$ is positive semidefinite and as, in the previous point,  entanglement cannot be generated.\\

\noindent
$3.$\quad The two qubits couple to a same set of fields: then, 
$\phi^1_j=\phi^2_j=\phi_j$, $c^1_{jk}=c^2_{jk}=c_{jk}$ and the matrix $D$ reduces to the form $D=\begin{pmatrix}\Delta&\Delta\cr \Delta&\Delta\end{pmatrix}$ where $\Delta$ is a $3\times 3$ matrix with entries 
\begin{equation}
\label{mat1}
\Delta_{jk}=\sum_\ell\Big\{\Re( \overline{c_{j\ell}}\, c_{k\ell})\,\coth\left(\frac{\beta\omega_\ell}{2}\right)+i\,\Im( \overline{c_{j\ell}}\, c_{k\ell})\Big\}\,.
\end{equation}
By choosing $\vert u\rangle=\vert v\rangle$ in~\eqref{entgenNM}, namely $\vert\psi\rangle=\vert\phi_\perp\rangle$, the (strict) inequality becomes
\begin{equation}
\label{cond2}
\left|\langle u\vert\Im(\Delta)\vert u\rangle\right|>0\ ,
\end{equation}
where
\begin{equation}
\label{D3}
\Im(\Delta)=\frac{\Delta-\Delta^T}{2i}=\frac{1}{2i}\begin{pmatrix}
0&x_{12}&x_{13}\cr
-x_{12}&0&x_{23}\cr
-x_{13}&-x_{23}&0
\end{pmatrix}\ ,
\end{equation}
with $x_{jk}=\Im(\langle C_k\vert C_j\rangle)$ with $\vert C_j\rangle$ the vector of components $c_{j\ell}$. Then one computes
\begin{equation}
\label{entgensimp}
\langle u\vert\Im(\Delta)\vert u\rangle=\sum_{j,k=1}^3x_{jk}\,\Im(\overline{u_j}\,u_k)\ .
\end{equation}
If for instance $x_{12}\neq 0$, inequality~\eqref{cond2} is satisfied by
choosing $\vert\psi\rangle$ such that $\sigma_3\vert\psi\rangle=\vert\psi\rangle$. In such a case $\vert u\rangle=(1,-i,0)$ and $\langle u\vert\Im(\Delta)\vert u\rangle=-1$ so that the separable state $\vert\psi\rangle\otimes\vert\psi_\perp\rangle$ becomes entangled.\\

\subsection{Stochastic derivation}

Non-Markovian dynamical maps different from those  microscopically derived as in the previous Section result from from an alternative approach where one chooses an effective description of the environment.

It has been recently shown~\cite{DioFer14} that, if the correlation functions $\mathfrak{D}_{jk}^{\alpha\gamma}(\tau,s;t_0):=D_{jk}^{\alpha\gamma}(\tau,s)$ are inserted into the map~\eqref{GMtt0b}, then it is unravelled by the following stochastic Schr\"odinger equation:
\begin{eqnarray}
\label{SSEdiss}
\hskip-.7cm
\frac{d|\psi_t\rangle}{dt}&=&-i\sum_{j,k=1}^3\sum_{\alpha,\gamma=1}^2\sigma^{\alpha}_j(t)\bigg(\phi_j^\alpha(t)+\int_0^tds 
[\overline{D^{\alpha\gamma}_{jk}}(t,s)-S^{\alpha\gamma}_{jk}(t,s)]\frac{\delta}{\delta\phi_k^\gamma}(s)\bigg)|\psi_t\rangle\ ,
\label{SSEdiss2}
\end{eqnarray}
where $\delta/\delta\phi_k^\gamma(s)$ denotes a functional derivative, and the stochastic correlation functions are defined in~\eqref{corrD}-\eqref{corrS}. This means that the average dynamics of an initial state $\vert\psi_{t_0}\rangle$ provided by this equation recovers $\Mc_{t,t_0}$, i.e.
\begin{equation}
\mathbb{E}\left[|\psi_t\rangle\langle\psi_t|\right]=\Mc_t[|\psi_{t_0}\rangle\langle\psi_{t_0}|]\,.
\end{equation}
Notice that Eq.~\eqref{GMtt0b} does not contain $S$: this reflects the fact that there is an infinite number of stochastic Schr\"odinger equations unraveling the same map. Furthermore, none of them can be reduced to a Hamiltonian coupling with classical stochastic fields, unless they are real~\cite{DioFer14}, in which case, the term in~\eqref{SSEdiss2} with the functional derivative disappears.

We now show that, unlike microscopically derived non-Markovian dynamics where the entanglement generation properties do not depend on the initial time $t_0\geq 0$, the possibilities offered by the 
stochastic derivation are larger in that the entanglement generation properties may change in time.

Such a different behaviour can be explained as follows:  in the Markovian regime each stochastically derived Gaussian open dynamics also admits a microscopic derivation. Indeed, as previously mentioned, time-translation invariant correlation functions can be obtained from coupling to suitable microscopic thermal environments in equilibrium that include those with Dirac delta correlation functions that result in Markovian open dynamics.
On the other hand, this is no longer true in the non-Markovian regime, where there may exist stochastically generated dissipative Gaussian dynamics characterized by correlation functions that do not obey the time dependence in~\eqref{boscorrt0}, \textit{i.e.} which cannot be obtained from coupling to bosonic Gaussian baths. As a consequence, if the reduced dynamical maps $\Mc_{t,t_0}$ of Eq.~\eqref{GMtt0b} involve two-point correlation functions with time-dependence not of the form 
$D(t-t_0,s-t_0)$, the maps do not in general depend just on $t-t_0$ only and may display, depending on $t_0$, different entangling generation properties compared with $\Mc_{t,0}$. These properties are determined by Eq.~\eqref{entgenNM} where now $D^{\alpha\gamma}_{jk}$ have an explicit dependence on $t_0$.

An example of such a possibility follows from considering a map $\Mc_{t,0}$ of stochastic origin describing two qubits coupled to a same set of complex stochastic fields 
\begin{equation}
\phi_j(t)=\sum_{\ell=1}^3\mu_{j\ell}W_\ell(t)\,+\,c_j\ ,
\end{equation}
where $\mu_{j\ell}$ are suitable complex coefficients and $W_\ell(t)$ are Wiener processes with zero mean and two-point correlation functions 
$\mathbb{E}[W_j(t)W_k(s)]=\delta_{jk}\,\min(t,s)$, while $c_j$ are deterministic values of the fields at $t=0$.  Then, 
\begin{equation}
D^{\alpha\gamma}_{jk}(t,s)=\min(t,s)\,\sum_{\ell=1}^3 \mu_{j\ell}\overline{\mu_{k\ell}}\,+\,c_j\,\overline{c_k}\ ,
\end{equation}
for all $\alpha,\gamma=1,2$, whence, at $t_0\geq 0$ one derives a Kossakowski matrix $D(t_0)=\begin{pmatrix}\Delta(t_0)&\Delta(t_0)\cr \Delta(t_0)&\Delta(t_0)\end{pmatrix}$ where $\Delta(t_0)$ has entries
\begin{equation}
\label{stochent}
\Delta_{jk}(t_0)=t_0\,\sum_{\ell=1}^3 \mu_{j\ell}\overline{\mu_{k\ell}}\,+\, c_j\,\overline{c_k}\ .
\end{equation} 
The entanglement generation criterion~\eqref{cond2} can then be applied to the matrix $\Im(\Delta(t_0))$ with entries 
\begin{equation}
t_0\,\sum_{\ell=1}^3\Im( \mu_{j\ell} \overline{\mu_{k\ell}})+\Im( c_j\,\overline{c_k})\ ,
\end{equation}
and explicitly reads
\begin{align}
|\langle u\vert\Im(\Delta(t_0))\vert u\rangle|=\frac{1}{2}&\Bigg| t_0 \sum_{\ell=1}^3 \Bigg(  \Big| \sum_{j} \overline{u_j} \mu_{j\ell}\Big|^2 -  \Big|\sum_{j} u_j \mu_{j\ell}\Big|^2  \Bigg) +\Bigg(  \Big|\sum_{j} \overline{u_j} c_j \Big|^2 - \Big|\sum_{j} u_j c_j \Big|^2 \Bigg) \Bigg| >0 .
\end{align}
Evidently, one can choose the initial fields and the complex coefficients $\mu_{j\ell}$ 
such that $\Im(\Delta(0))=0$ while $\langle u\vert\Im(\Delta(t_0))\vert u\rangle\neq 0$ so that 
$\vert\psi\rangle\otimes\vert\psi_\perp\rangle$ cannot be entangled at $t=0$, but becomes entangled at any $t_0>0$. 
Vice versa, one can arrange the coefficients in order to ensure that $\Im(\Delta(0))\neq0$, while 
$\Im(\Delta(t_0))=0$ at some $t_0$ so that there is entanglement generation at $t=0$ but not at $t_0>0$.

Finally, notice that if the classical Gaussian fields are real, there is no entanglement generation, since then the stochastic Schr\"odinger equation~\eqref{SSEdiss} becomes of Hamiltonian form, describing two independent qubits interacting with a classical background. Consequently, non-local correlations between non-interacting qubits cannot be created solely by classical means.

\section{Markov vs Non-Markov}
In the previous section we have seen that the criteria for entanglement generation  
in the Markovian [Eq.~\eqref{entgenM}] and non-Markovian [Eq.~\eqref{entgenNM}] regimes  
have a similar structure. It is thus important to discuss their differences.

\subsection{Time dependence}
A first important difference between Markovian and non-Markovian entanglement generation is the time dependence. In the previous section we have indeed seen that in non-Markovian dynamics the first order of the expansion of the map $\Mc_t$ (see~\eqref{expansNM}) for small positive times vanishes, while it 
it does not in the Markovian case. As already remarked, this fact is strictly related to the singular delta-like correlation functions of Markovian baths. Entanglement is generated on a time scale (say $\tau_G$), such that the approximation given by the Dyson expansion is valid, that is  $0< \tau_G<t^2$ for non-Markovian dynamics, and $0<\tau_G<t$ for Markovian ones. Accordingly, in the Markovian regime and for small times entanglement generation occurs on a longer time-scale ($\propto t$) than in 
the non-Markovian regime ($\propto t^2$).

In order to better illustrate the role of the difference in the time-dependence in the two regimes, let us consider a \lq pure dephasing\rq~model of one qubit interacting with a stochastic real Ornstein-Uhlenbeck process through a Hamiltonian 
\begin{eqnarray}
\label{deph}
H(t)&=&\omega_z \sigma_z\,+\,\sigma_z \phi(t)\\ 
D_\epsilon(\tau,s)&=&\mathbb{E}[\phi(\tau)\phi(s)]
=\frac{1}{2\epsilon}\exp\Big(-\frac{ |\tau-s|}{\epsilon}\Big)\ .
\label{OU}
\end{eqnarray}
Since $H_S$ and $H_I$ commute, taking the time-derivative of $\Mc_t$ in~\eqref{GMta} and~\eqref{GMtt0b} yields the following master equation in the interaction picture:
\begin{equation}
\label{mepd}
\dot{\rho}_t=-\left(\int_0^t D_\epsilon(t-s)\,ds\right)\, \left[\sigma_z,\left[\sigma_z,\rho_t\right]\right]\,.
\end{equation}
By integrating this equation as a Dyson series and stopping at the first order, one obtains  
\begin{eqnarray}
\label{ropd}
\hskip -.5cm
\rho_t&\simeq&\rho-\left(\int_0^t d\tau\int_0^{\tau}D_\epsilon(\tau-s)\,ds\right)\left[\sigma_z,\left[\sigma_z,\rho\right]\right]\\
\hskip -.7cm
&=&\rho-\left(\frac{t}{2}+\frac{\epsilon}{2}\left(e^{- t/\epsilon}-1\right)\right) \left[\sigma_z,\left[\sigma_z,\rho\right]\right]\, .
\label{ropdexp}
\end{eqnarray}
Notice that the bath correlation function approximates the Dirac delta when $\epsilon\rightarrow0$; one can thus recover a Markovian dynamics in that limit. Moreover,
replacing $D_\epsilon(t-s)$ by $\delta(t-s)$ in Eq.~\eqref{ropd} yields the same result as taking the limit $\epsilon\rightarrow0$ in Eq.~\eqref{ropdexp}:
\begin{equation}
\label{Dyson}
\rho_t\simeq\rho-\frac{t}{2}\ \left[\sigma_z,\left[\sigma_z,\rho\right]\right]\, .
\end{equation}
Instead, performing a short time expansion in~\eqref{expansNM} or in~\eqref{ropdexp}  one obtains
\begin{equation}\label{smalltimeexp}
\rho_t\simeq\rho-\frac{ t^2}{4\epsilon} \left[\sigma_z,\left[\sigma_z,\rho\right]\right]\,.
\end{equation}
While in the Markovian regime the first non-trivial contribution to the Dyson expansion is of order $t$ and independent of $\epsilon$ [Eq.~\eqref{Dyson}], the small time expansion of Eq.~\eqref{smalltimeexp} is of order $t^2$ and diverges when $\epsilon\rightarrow0$. This indicates that the Markovian limit $\epsilon\rightarrow0$ cannot be exchanged with the small time expansion: indeed the small parameter $t^2/\epsilon$ becomes large with small $\epsilon$.
Therefore, while the Markovian regime can be derived as a singular limit of the non-Markovian one, nevertheless one cannot obtain the Markovian entanglement criterion as a limit case of the non-Markovian one.

\subsection{Entanglement criterion}
In this section we investigate in more detail the phenomenon presented in the previous Section, namely that the entanglement generation properties in the Markovian regime cannot be obtained from the non-Markovian ones through a continuous family of approximants of the Dirac delta. In order to do so, we consider the following family of functions parametrized by $\epsilon$:
\begin{equation}\label{deps}
d_\epsilon(t):=\epsilon a\left(\frac{t}{\epsilon}\right) + b\left(\frac{t}{\epsilon}\right) +\frac{1}{\epsilon}c\left(\frac{t}{\epsilon}\right)\,,
\end{equation}
where $a(t)$, $b(t)$, $c(t)$ are integrable, continuous functions, and $a(0),b(0)\neq 0$. We further assume that $c$ is an approximation of the Dirac delta, i.e. that in the limit $\epsilon\rightarrow 0$, $c(t/\epsilon)/\epsilon\rightarrow\delta(t)$. One can readily promote this description to the matrix formalism by assuming the same structure as in~\eqref{deps} for all correlation functions $D^{\alpha\gamma}_{jk}(\tau-s)$.

Correlation functions of this kind provide different entanglement criteria in the two regimes.  
Indeed, in the non-Markovian case the criterion~\eqref{entgenNM} is based on the correlation function at $t=0$, 
\begin{equation}
\label{MMcrr}
d_\epsilon\left(0\right)=\epsilon a\left(0\right) + b\left(0\right) +\frac{1}{\epsilon}c\left(0\right)\ .
\end{equation} 
Therefore, the entanglement generation properties depend on the three functions $a(t)$, $b(t)$ and $c(t)$.

On the other hand, taking the Markovian limit $\epsilon\rightarrow 0$, because of the assumed integrability of the functions, they vanish at infinity, whence the contribution from $a$ is suppressed, while the one from $b$ is hidden by the divergence of $1/\epsilon$. Then, the entanglement generation properties in the Markovian regime are solely determined by the function $c(t)$.
Accordingly, Eq.~\eqref{deps} displays a family of correlation functions providing different entanglement generation properties in the Markovian and non-Markovian regimes.

In the following we investigate whether it is possible to build correlation functions exhibiting the above behaviour within the models considered in the previous sections.\\
Given a microscopic model of two qubits interacting with a bosonic thermal bath with correlation functions as in~\eqref{Dgen}, in order to achieve a Markovian behaviour, the spectral densities $J^{\alpha\gamma}_{jk}(\omega)$ must ultimately provide Dirac deltas in time and this can only be achieved by a Ohmic behaviour ($\propto\omega$), and by a sufficiently low inverse temperature $\beta$ such that $\coth\left(\omega\beta/2\right)\simeq 2/\omega\beta$. Under these conditions, we have that the real parts of the correlation functions in the criterion~\eqref{entgenNM} behave as in~\eqref{deps} with 
$a=b=0$~\cite{BrePet02}. This implies that, in a model of this kind, if entanglement generation does (does not) occur  in the non-Markovian regime, it also does (does not) in the Markovian one.\\
On the other hand, if one considers an effective stochastic description the situation may be different,  thus offering further evidence that, though the two approaches are equivalent in the Markovian regime~\cite{BarGre09}, instead deriving non-Markovian dynamics by stochastic means provides a richer scenario than by a microscopic approach. 
Indeed, consider two qubits interacting with a same stochastic field
\begin{equation}
\phi_\epsilon\left(\frac{t}{\epsilon}\right)=\sqrt{\epsilon} \phi_1\left(\frac{t}{\epsilon}\right) + \frac{1}{\sqrt{\epsilon}}\phi_2\left(\frac{t}{\epsilon}\right)\,,
\end{equation}
where $\phi_1$ and $\phi_2$ are classical complex stochastic fields where $\phi_2$ tends to a white noise when $\epsilon$ goes to zero (e.g. the Ornstein-Uhlenbeck process of Eq.~\eqref{OU}). One can easily check that the two-point correlation functions $D_{jk}(\tau,s)$ can be recast in the form~\eqref{deps}, where
\begin{eqnarray}
a\left(\frac{t}{\epsilon}\right)&=&\mathbb{E}\left[\phi_1^*\left(\frac{t}{\epsilon}\right)\phi_1\right]\,,\\
b\left(\frac{t}{\epsilon}\right)&=&\mathbb{E}\left[\phi_1^*\left(\frac{t}{\epsilon}\right)\phi_2+\phi_2^*\left(\frac{t}{\epsilon}\right)\phi_1\right]\,,\\
c\left(\frac{t}{\epsilon}\right)&=&\mathbb{E}\left[\phi_2^*\left(\frac{t}{\epsilon}\right)\phi_2\right]\,.
\end{eqnarray}
Therefore, by an appropriate choice of the stochastic fields $\phi_{1,2}$ and thus of $a(0)$, $b(0)$, this kind of model may generate entanglement in the non-Markovian regime even if, the parameters of the type $c(0)$ are not able to guarantee it in the Markovian one. It may also happen that the choice of $c(0)$ enforces entanglement generation in the Markovian regime, when $\epsilon\to0$, but that of $a(0)$ and $b(0)$ could forbid it for large $\epsilon$.

\section{Conclusions}
We have investigated the entanglement generation properties of Gaussian non-Markovian dynamics at initial time $t=t_0$. We have shown that the entanglement generation criterion has the same structure as in the Markovian regime, provided that the Kossakowski matrix is replaced by the matrix whose entries are two-point bath correlation functions at $t=t_0$. Moreover, in the non-Markovian regime, Hamiltonian terms do not contribute to entanglement generation.
Although one recovers the Markovian regime from the non-Markovian one when the two-point correlation functions become singular, this is not true for the entanglement generation conditions;  furthermore, in the Markovian regime entanglement generation occurs on a longer time-scale 
($\propto t$) than in the non-Markovian regime  where the time-scale is $\propto t^2$.
Finally, if the correlation functions are time-translation invariant, the entangling properties of the dynamics do not change if the initial qubit state is set, not at $t=0$, but at $t_0>0$.
On the other hand, in absence of time-translation invariance, we showed that the microscopic and stochastic derivations of Gaussian open dynamics, which are equivalent in the Markovian regime may differ in the non-Markovian one.
Indeed, we proved that two qubit open dynamics derived from couplings to generic Gaussian bosonic baths, even out of equilibrium, exhibit entanglement generation properties that cannot vary in time. On the other hand, suitable non-time translation invariant couplings of the two qubits to Gaussian stochastic fields may in general provide
entanglement generation properties that may depend on time.

\section*{Acknowledgements} 
The authors thank R. Floreanini and G. Gasbarri for useful discussions.
 The work of LF was supported by the TALENTS$^3$ Fellowship Programme (CUP code J26D15000050009, FP code 1532453001, managed by AREA Science Park through the European Social Fund), and by the Royal Society under the Newton International Fellowship No NF170345.


\end{document}